\documentclass[11pt]{article}

\usepackage{euscript}
\usepackage{amssymb}
\usepackage{amsfonts}
\usepackage{amsbsy}
\usepackage{amsmath}
\usepackage{epsfig}
\usepackage{amsthm}
\usepackage{amscd}
\usepackage{amstext}
\usepackage{verbatim}

\usepackage[pdftex]{hyperref}

\def\ben{\begin{equation}}
\def\een{\end{equation}}

\let\a=\alpha \let\b=\beta  \let\d=\delta 
   \let\k=\kappa
     
\let\s=\sigma

\let\w=\omega  \let\D=\Delta

\def\be{\begin{equation}}
\def\ee{\end{equation}}
\def\beq{\begin{equation}}
\def\eeq{\end{equation}}
\def\ba{\begin{array}}
\def\ea{\end{array}}

\def\dalemb#1#2{{\vbox{\hrule height .#2pt
       \hbox{\vrule width.#2pt height#1pt \kern#1pt
               \vrule width.#2pt}
       \hrule height.#2pt}}}

\newcommand{\bea}{\begin{eqnarray}}
\newcommand{\eea}{\end{eqnarray}}

\def\Lag{{\mathcal{L}}}

\def\ocal{{\mathcal{O}}}

\begin{document}

\begin{center}

{ \LARGE {\bf Electron star birth:\\
A continuous phase transition at nonzero density}}

\vspace{1cm}

{\large Sean A. Hartnoll and Pavel Petrov

\vspace{0.7cm}

{\it Center for the Fundamental Laws of Nature, \\
Department of Physics, Harvard University,\\
Cambridge, MA 02138, USA \\} }

\vspace{1.6cm}

\end{center}

\begin{abstract}

We show that charged black holes in Anti-de Sitter spacetime can undergo a third order phase transition at a critical temperature
in the presence of charged fermions. In the low temperature phase, a fraction of the charge is carried by a fermion fluid located a finite distance from the black hole. In the zero temperature limit the black hole is no longer present and all charge is sourced by the fermions. The solutions exhibit the low temperature entropy density scaling $s \sim T^{2/z}$ anticipated from the emergent IR criticality of recently discussed electron stars.

\end{abstract}

\pagebreak
\setcounter{page}{1}

Recent work has argued that `electron stars', a planar fluid of charged fermions in Anti-de Sitter space in equilibrium under gravitational and electromagnetic forces, provide a compelling holographic framework in which to study metallic quantum criticality \cite{Hartnoll:2009ns, Hartnoll:2010gu, Hartnoll:2010xj}. Two key features of electron star solutions are emergent criticality at low energy with a finite dynamical critical exponent  \cite{Hartnoll:2009ns, Hartnoll:2010gu} and the presence of a `smeared' Fermi surface \cite{Hartnoll:2010xj}. Metallic criticality is difficult to study with conventional field theoretic techniques as the many gapless excitations of the Fermi surface cannot be integrated out and must be included in a strongly interacting IR fixed point \cite{review}.

The zero temperature electron stars studied in \cite{Hartnoll:2010gu} are also interesting as charged `solitonic' gravitational configurations without a black hole horizon. Their existence is aided by the gravitational well of the asymptotically Anti-de Sitter spacetime (AdS). In this sense they are the fermionic analogues of zero temperature holographic superconductors \cite{Hartnoll:2008kx, Gubser:2009cg, Horowitz:2009ij}. At sufficiently high temperatures in the holographically dual nonzero density field theory we might expect the electron star to collapse and form a Reissner-Nordstrom black hole. This would be analogous to the fact that zero temperature stars in AdS with spherical, rather than planar, symmetry undergo a first order transition to a black hole as a function of the energy of the star in units of the size of the spatial sphere at the boundary \cite{deBoer:2009wk, Arsiwalla:2010bt}. The primary result of this letter is that electron stars undergo a third order transition to a charged black hole above a critical temperature determined by the chemical potential and mass of the fermions. Lowering the temperature through the transition, the birth of the electron star, thereby corresponds to a continuous phase transition in the nonzero density dual field theory characterised by the appearance of a smeared Fermi surface.

This letter will study the dynamics of a 3+1 dimensional ideal fluid of charged relativistic fermions coupled to electromagnetism and gravity with a negative cosmological constant. The Lagrangian may be written
\be\label{eq:lagrangian}
\Lag = \frac{1}{2 \k^2} \left(R + \frac{6}{L^2} \right) - \frac{1}{4 e^2} F_{ab} F^{ab} + p(\mu,s) \,.
\ee
This is the Schutz form of the action for a gravitating ideal fluid in terms of the fluid pressure $p$ \cite{Schutz:1970my}, generalised to allow  the fluid to be charged \cite{Hartnoll:2010gu}. The local chemical potential $\mu = |d \phi + \a d \b + \theta ds + A|$, where $\{\phi,\a,\b \}$ are fluid potential variables, $s$ the local entropy density and $\theta$ the thermasy. This action leads to the expected ideal fluid equations of motion as described in e.g. \cite{Schutz:1970my, Hartnoll:2010gu}.

The fluid Lagrangian (\ref{eq:lagrangian}) is a coarse grained description of the fermions in which the fermion physics is subsumed into locally defined thermodynamic quantities. In the context of astrophysics this could be called the Tolman-Oppenheimer-Volkoff description \cite{tolman, openn} while in condensed matter physics it is known as the Thomas-Fermi approximation \cite{thomas, fermi}. In our context, the fluid Lagrangian (\ref{eq:lagrangian}) is  a correct description of the system when \cite{Hartnoll:2010gu}
\be\label{eq:sim}
e^2 \sim \frac{\k}{L} \ll 1 \,. 
\ee
Without loss of generality we will take the fermions to have unit charge and mass $m$. For detailed discussions of the connection between microscopic and fluid descriptions of gravitating fermions, see e.g. \cite{Ruffini:1969qy, Arsiwalla:2010bt}.

The spacetime metric and Maxwell field take the general form
\be\label{eq:metric}
ds^2 = L^2 \left(- f dt^2 + g dr^2 + \frac{1}{r^2} \left( dx^2 + dy^2 \right) \right) \,, \qquad A = \frac{e L}{\k} h dt \,.
\ee
For the fermion fluid, a crucial role is played by the local chemical potential
\be\label{eq:mulocal}
\mu_\text{loc.} = A_{\hat t} = \frac{A_t}{L \sqrt{f}} = \frac{e}{\k} \frac{h}{\sqrt{f}} \,.
\ee
This chemical potential determines the local thermodynamic quantities of the fermion fluid. Before specifying these,
however, we need to discuss how a nonzero temperature is going to appear in our equations. Placing the dual field theory
in equilibrium at a nonzero temperature can be described by compactifying the Euclidean time direction. This implies that the
Euclidean time direction in the bulk must also be compactified. Two consequences of a periodic bulk Euclidean time circle
are that for regularity of the spacetime, recall that we are in planar coordinates, we must have a finite size black hole
horizon in the interior and secondly that the local fermion fluid equation of state is at finite temperature. Physically this describes
a black hole surrounded by a fermion fluid in thermal equilibrium with the Hawking radiation. In our bulk classical limit the effects of
Hawking radiation should be negligible, while the black hole remains present. Therefore, we should expect that in this limit
we can treat the fermions as a zero temperature fluid in a black hole background. We will make this statement precisely towards the end of the paper.

The upshot of the previous paragraph is that we can use the same zero temperature equations for the Einstein-Maxwell-fluid system as for zero temperature electron stars, but with boundary conditions ensuring that the spacetime has a finite temperature horizon. This means that 
the energy density, charge density and pressure of the fermion fluid are determined by the local potential (\ref{eq:mulocal}) via their zero temperature equation of state
\be\label{eq:eos}
\hat \rho = \hat \b \int_{\hat m}^{\frac{h}{\sqrt{f}}}  \epsilon^2 \sqrt{\epsilon^2 - \hat m^2} d\epsilon \,, \qquad \hat \sigma = \hat \b \int_{\hat m}^{\frac{h}{\sqrt{f}}}  \epsilon \sqrt{\epsilon^2 - \hat m^2} d\epsilon \,, \qquad - \hat p = \hat \rho - \frac{h}{\sqrt{f}} \hat \sigma \,. 
\ee
Here we have introduced dimensionless hatted quantities
\be
p = \frac{1}{L^2 \k^2} \hat p \,, \qquad \rho = \frac{1}{L^2 \k^2} \hat \rho \,,\qquad \sigma = \frac{1}{e L^2 \k} \hat \sigma \,, \qquad \hat \b = \frac{e^4 L^2}{\k^2} \b \,, \qquad \hat m^2 = \frac{\k^2}{e^2} m^2 \,.
\ee
Throughout we will use hats to denote bulk or field theory quantities that have been rescaled by factors of $\{e,L,\k\}$ in such a way that no such factors appear in the equations.
The constant $\b$ is a dimensionless order one number that counts the microscopic degrees of freedom of the fermion. The important fact about $\hat \b$ is that it measures the ratio of the Maxwell and Newton couplings and should also be order one if we are in the regime (\ref{eq:sim}).
Furthermore, the equations of motion following from the action (\ref{eq:lagrangian}) imply the same equations for the components of the fields as in \cite{Hartnoll:2010gu}
\bea
\frac{1}{r} \left(\frac{f'}{f} + \frac{g'}{g} + \frac{4}{r}\right) + \frac{g h \hat \sigma}{\sqrt{f}} & = & 0 \,, \label{eq:a}\\
\frac{f'}{r f} - \frac{h'^2}{2f} + g (3 + \hat p) - \frac{1}{r^2} & = & 0 \,, \\
h'' + \frac{g \hat \sigma}{\sqrt{f}} \left(\frac{r h h'}{2}  - f \right) & = & 0 \,. \label{eq:b}
\eea

From (\ref{eq:eos}) it is immediate that a nonvanishing density of fermions at a particular radius requires that the mass be lower than the local chemical potential $m < \mu_\text{loc.}$, or $\hat m < h/\sqrt{f}$. In the Reissner-Nordstrom black hole solution
\be\label{eq:RN}
f =  \frac{1}{r^2} - \left(\frac{1}{r_+^2} + \frac{\hat \mu^2}{2} \right) \frac{r}{r_+} + \frac{\hat \mu^2}{2} \frac{r^2}{r_+^2} \,, \qquad g = \frac{1}{r^4 f} \,, \qquad h = \hat \mu \left(1 - \frac{r}{r_+} \right) \,,
\ee
one can see that $h/\sqrt{f}$ is bounded with a maximum in between the horizon at $r=r_+$ and boundary at $r=0$. Thus if $\hat m$ is too large, the fluid density will vanish everywhere, $\hat \rho = \hat \s = \hat p = 0$, and Reissner-Nordstrom (\ref{eq:RN}) will be the exact solution. As the mass is lowered it will eventually equal the local chemical potential at a critical radius
\be\label{eq:conditions}
\left. \frac{h}{\sqrt{f}} \right|_{r=r_c} = \hat m \,, \qquad \left. \frac{d}{dr} \frac{h}{\sqrt{f}}\right|_{r=r_c} = 0 \,.
\ee
It is simple to solve these two equations. The two equations determine the critical radius at which the star is born, dimensionlessly expressed as $r_c/r_+$, and the critical temperature of the black hole over the rescaled chemical potential $T_C/\hat \mu$. Recall that the temperature of a Reissner-Nordstrom black hole is
\be\label{eq:temperature}
T = \frac{1}{4 \pi c} \left| \frac{df}{dr} \right|_{r=r_+} \,.
\ee
At this point $c=1$, we have included this factor for later convenience.

\begin{figure}[h]
\begin{center}
\includegraphics[width=210pt]{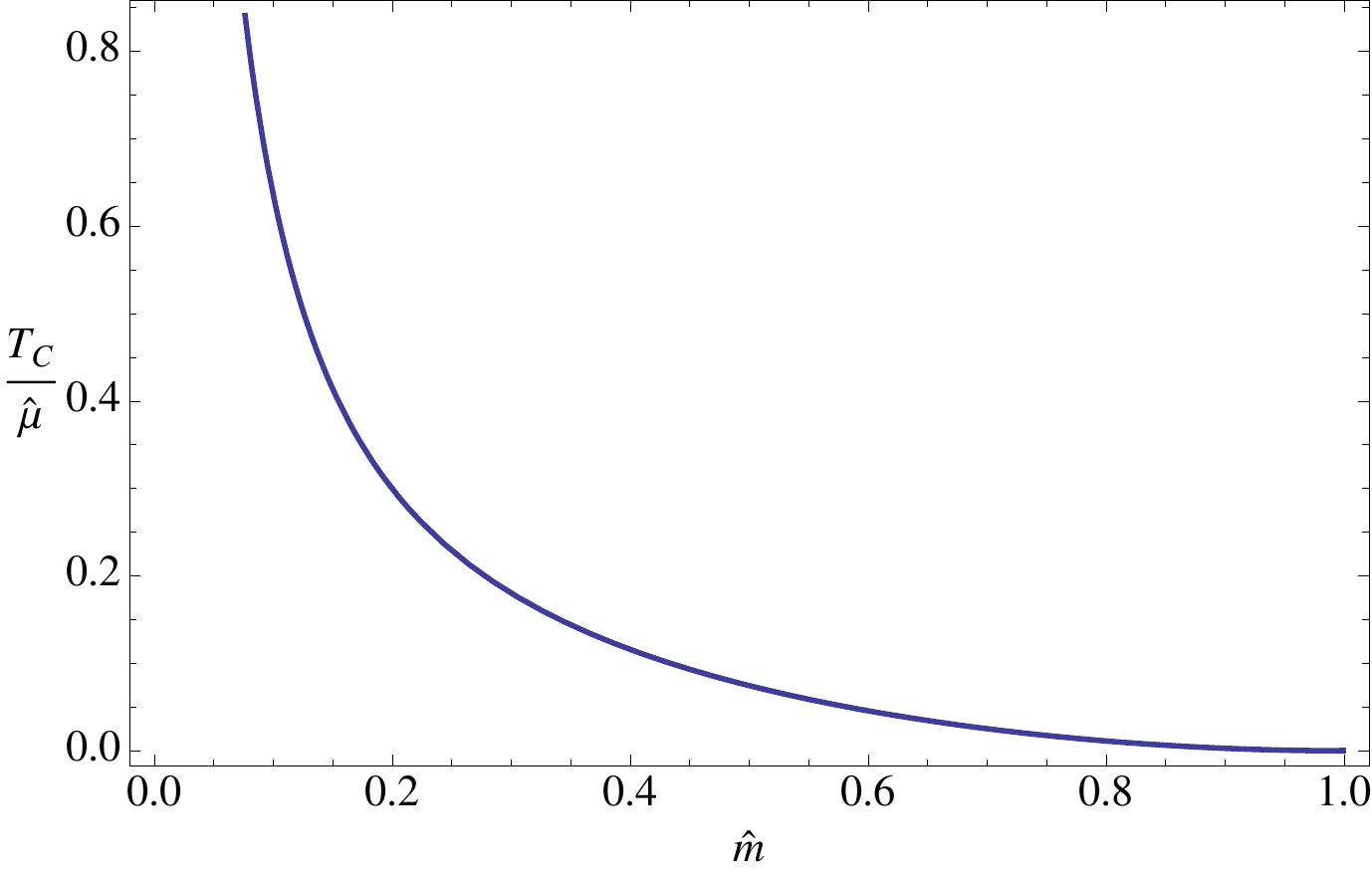}\hspace{0.1cm}\includegraphics[width=210pt]{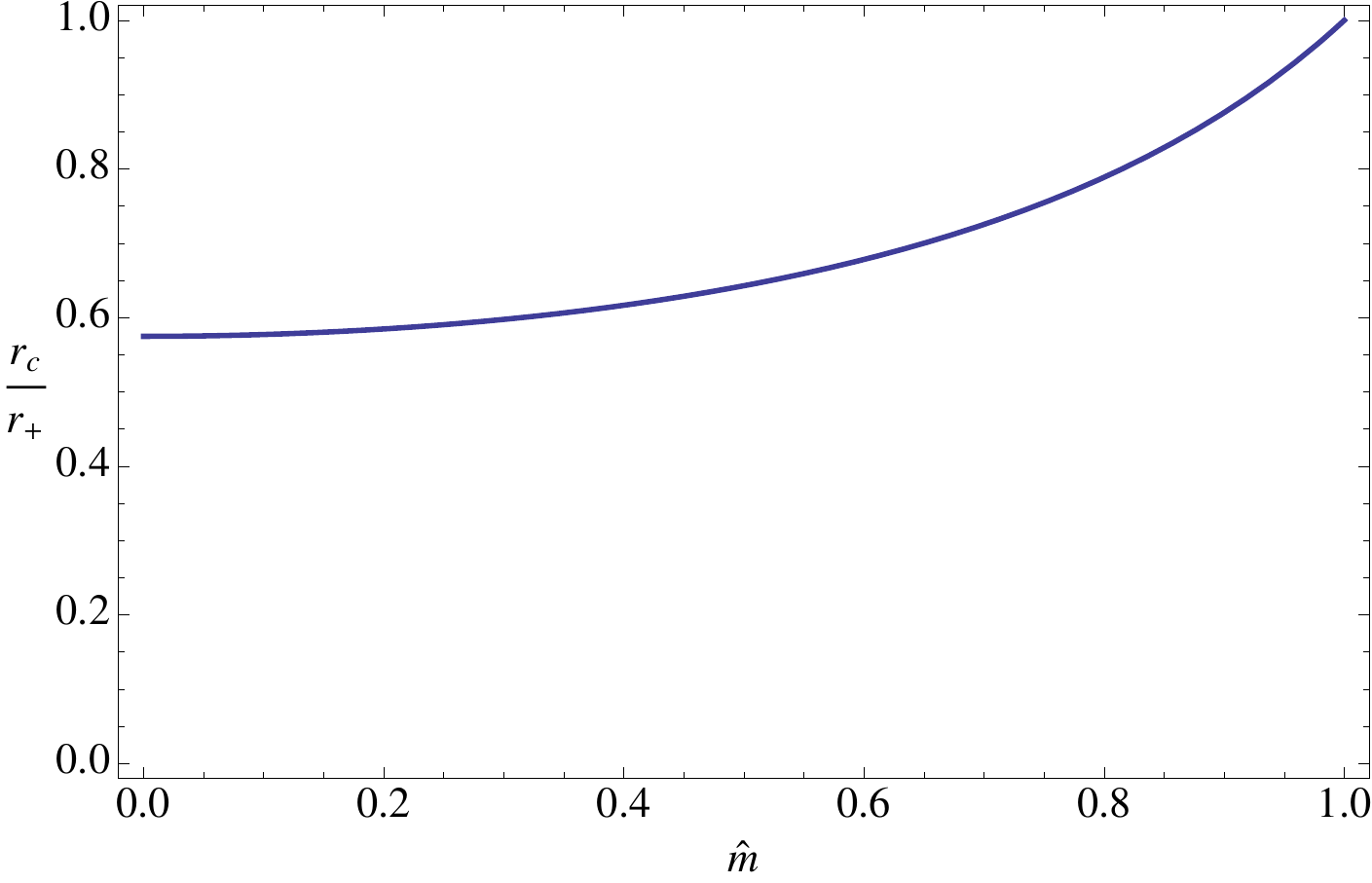}\caption{Critical temperature (left plot) and radius (right plot) at which the electron star is born, as a function of the fermion mass. These quantities do not depend on $\hat \beta$. \label{fig:critical}}
\end{center}
\end{figure}

In figure \ref{fig:critical} we plot the critical radius and critical temperature as a function of the fermion mass.
We see that the electron star forms at increasing low temperatures, and closer to the black hole horizon, as the fermion mass is increased. When the mass is increased beyond $\hat m = 1$ the star is never formed. This is consistent with the observation in \cite{Hartnoll:2010gu} that $\hat m < 1$ is necessary for the zero temperature electron star to exist. The critical temperatures and radii plotted in figure \ref{fig:critical} follow from the Reissner-Nordstrom black hole geometry and therefore do not depend on the parameter $\hat \beta$ appearing in the fluid equation of state (\ref{eq:eos}).
In fact, the conditions (\ref{eq:conditions}) simply characterise the radius at which a charged point particle can remain stationary, with gravitational and electromagnetic forces balancing.

Cooling below the critical temperature of figure \ref{fig:critical} we will find two solutions $\{r_1, r_2\}$ to the equation $\mu_\text{loc.} = m$ defining the boundary of the star. Thus the electron star broadens into a thin shell. Once broadened, the equation of state of the fluid as well as the backreaction of the fluid onto the geometry will be important. The solution at $T < T_C$ has three components:
\begin{enumerate}

\item Inner region, $r > r_2$. The solution has the Reissner-Nordstrom form (\ref{eq:RN}), but with $\hat \mu \to \hat \mu_0$, not directly related to the chemical potential of the dual field theory. This region determines the temperature through (\ref{eq:temperature}).

\item Intermediate region, $r_2 > r > r_1$. Here we must solve the equations (\ref{eq:a}) to (\ref{eq:b}) with $\hat \rho, \hat \sigma, \hat p$ nonzero. The quantities $\{f,g,h,h'\}$ must be matched onto the inner and exterior regions at $r_2$ and $r_1$.

\item Exterior region, $r_1 > r$. The solution again has the Reissner-Nordstrom form, although now it should be written as
\be\label{eq:RN2}
f =  c^2 \left(\frac{1}{r^2} - \hat M r + \frac{r^2 \hat Q^2}{2} \right) \,, \qquad g = \frac{c^2}{r^4 f} \,, \qquad h = c \left( \hat \mu - r \hat Q \right) \,.
\ee
The factor $c$, determining the normalisation of time, must be included in the definition of the temperature (\ref{eq:temperature}). Note the definitions are slightly modified with respect to \cite{Hartnoll:2010gu}.
\end{enumerate}

\begin{figure}[h]
\begin{center}
\includegraphics[width=210pt]{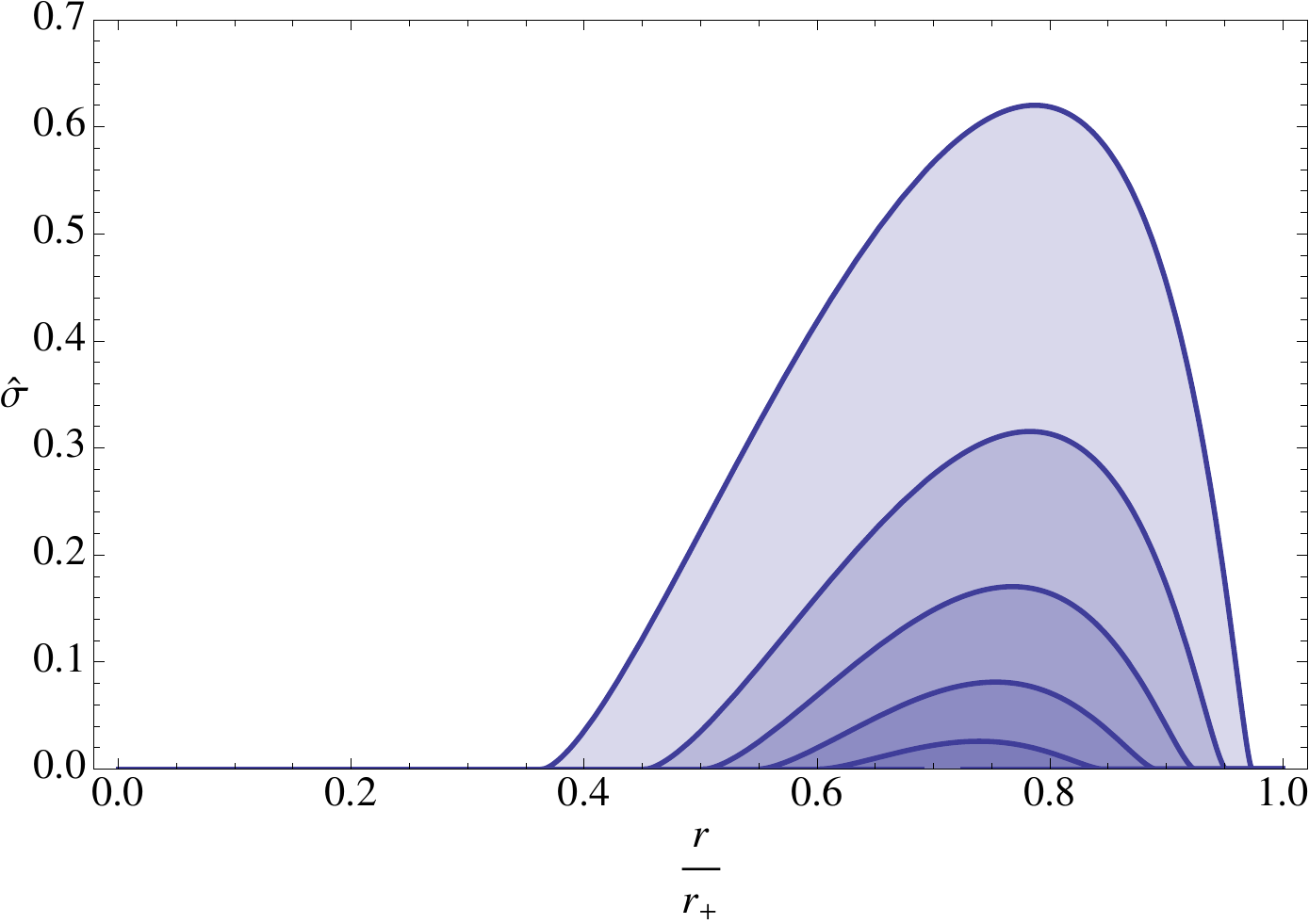}\hspace{0.1cm}\includegraphics[width=210pt]{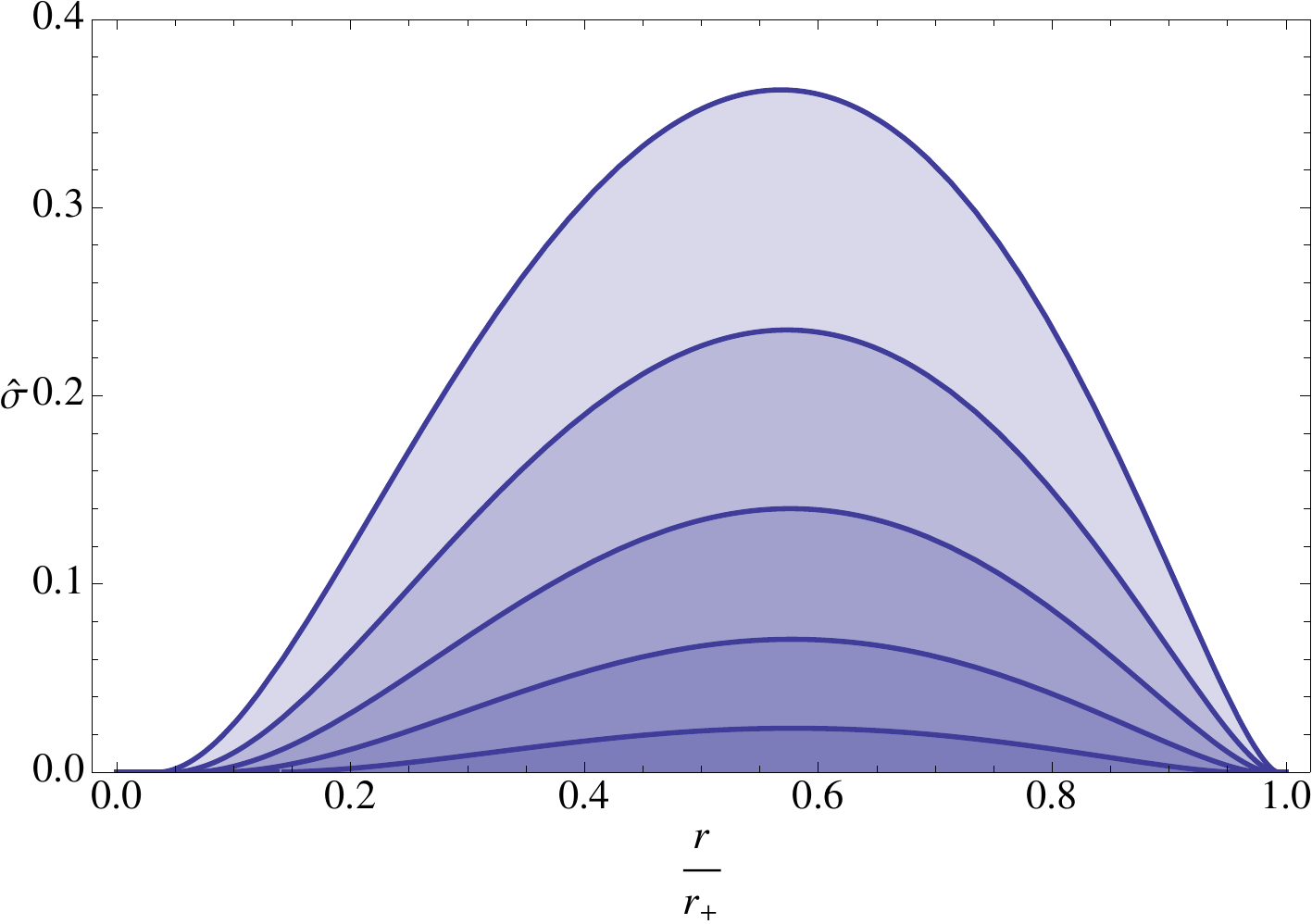}\caption{Radial density profiles of electron stars as a function of temperature. The curves show five temperatures between $0.07 \, T_C$ and $T_C$, with $\hat \mu$ held fixed. Both plots have $\hat \beta = 10$. The left plots has mass $\hat m = 0.7$ while the right plot has $\hat m = 0.1$. \label{fig:buildup}}
\end{center}
\end{figure}

In performing numerics one may without loss of generality set $r_+=1$. The solutions are then parametrised by $\hat \mu_0$ in the inner region. Via matching at $r_2$ and $r_1$, this initial condition then integrates forward to determine the values of the physical dual field theory quantities $\{c,\hat M, \hat Q, \hat \mu\}$. The solutions can then be labelled by the physical dimensionless ratio $T/\hat \mu$ or alternatively $T/T_C$. Performing the numerics is simple and Figure \ref{fig:buildup} shows two examples of how the density of fermions $\hat \sigma$ builds up and expands to widening shells as the temperature is lowered below the critical value.
In reading these plots one should keep in mind that the total charge in the fermion fluid will be \cite{Hartnoll:2010gu}
\be
\hat Q_\text{fermion}  =  \int^{r_2}_{r_1} \frac{\sqrt{g(s)}}{s^2} \hat \s(s) \, ds \,.
\ee
In particular, it is necessary to include the spatial volume element $\sqrt{g}/r^2$ in summing up the contributions from different radii to the total charge in the fermion fluid.

\begin{figure}[h]
\begin{center}
\includegraphics[width=250pt]{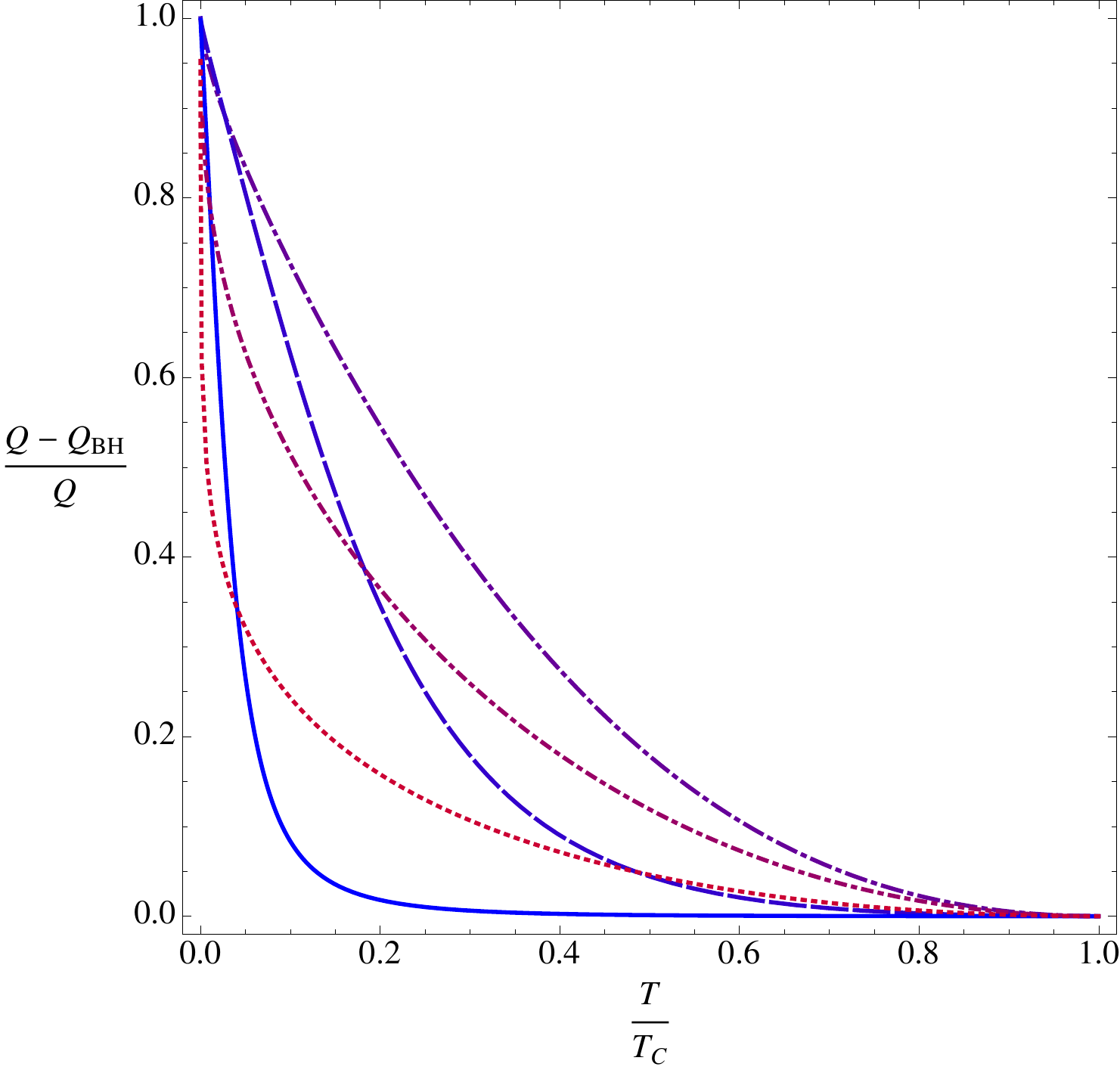}\caption{The fraction of charge carried by the fermion fluid as a function of temperature. All curves have $\hat \b = 10$ while from blue to red (solid to dotted) $\hat m = \{0.01, 0.07, 0.3, 0.55 ,0.75\}$. In quoting $T/T_c$, the chemical potential is kept fixed. \label{fig:Qratio}}
\end{center}
\end{figure}

The total charge density of the boundary field theory, read off from the exterior spacetime solution (\ref{eq:RN2}),  may be written
\be
\hat Q = \hat Q_\text{BH} + \hat Q_\text{fermion} \,,
\ee
where the charge density carried by the black hole is read off from the interior solution as
\be
\hat Q_\text{BH} = \frac{\hat \mu_0}{r_+} \,.
\ee
From the bulk perspective, the presence of the fermionic charge density is the defining characteristic of the electron star. This suggests examining the temperature dependence of the fraction of charge carried by the fermionic fluid, $(\hat Q - \hat Q_\text{BH})/\hat Q$, as a sort of bulk order parameter. The ratio will be zero for temperatures above the critical temperature and, based on the observation in \cite{Hartnoll:2010gu} that at zero temperature all the charge is carried by fermions, should tend to unity at low temperatures. Figure \ref{fig:Qratio} shows precisely this phenomenon. For both the low mass and heavier fermions, the charge is transferred from the interior black hole to the fermion fluid more slowly as the temperature is lowered than for intermediate mass fermions.

So far we have seen how the electron star emerges continuously from the Reissner-Nordstrom black hole solution as the temperature is lowered below a critical value. The electron star is characterised by the fact that a nonzero fraction of the charge is carried by a fluid outside of the black hole. These two observations suggest that the electron star birth is a second or higher order phase transition in the system. To make this claim precise, we must compute the behaviour of the free energy across the transition. The free energy density of the theory is most easily obtained from the thermodynamic relation
\be\label{eq:thermo}
\hat \Omega = \hat M - \hat \mu \hat Q - \hat s T \,.
\ee
Here $\hat s$ is the (rescaled) Bekenstein-Hawking entropy density given by
\be
\frac{L^2}{\k^2} \hat s = s = \frac{8 \pi}{\k^2} \frac{L^2}{4 r_+^2} \,. 
\ee
It was shown in \cite{Hartnoll:2010gu} that (\ref{eq:thermo}) held for zero temperature electron stars by showing that the Lagrangian (\ref{eq:lagrangian}) was a total derivative on shell. In the presence of a nonzero temperature horizon, it is easy to check that there is an additional contribution to the on shell action at the horizon that contributes the necessary $\hat s T$ term.

\begin{figure}[h]
\begin{center}
\includegraphics[height=250pt]{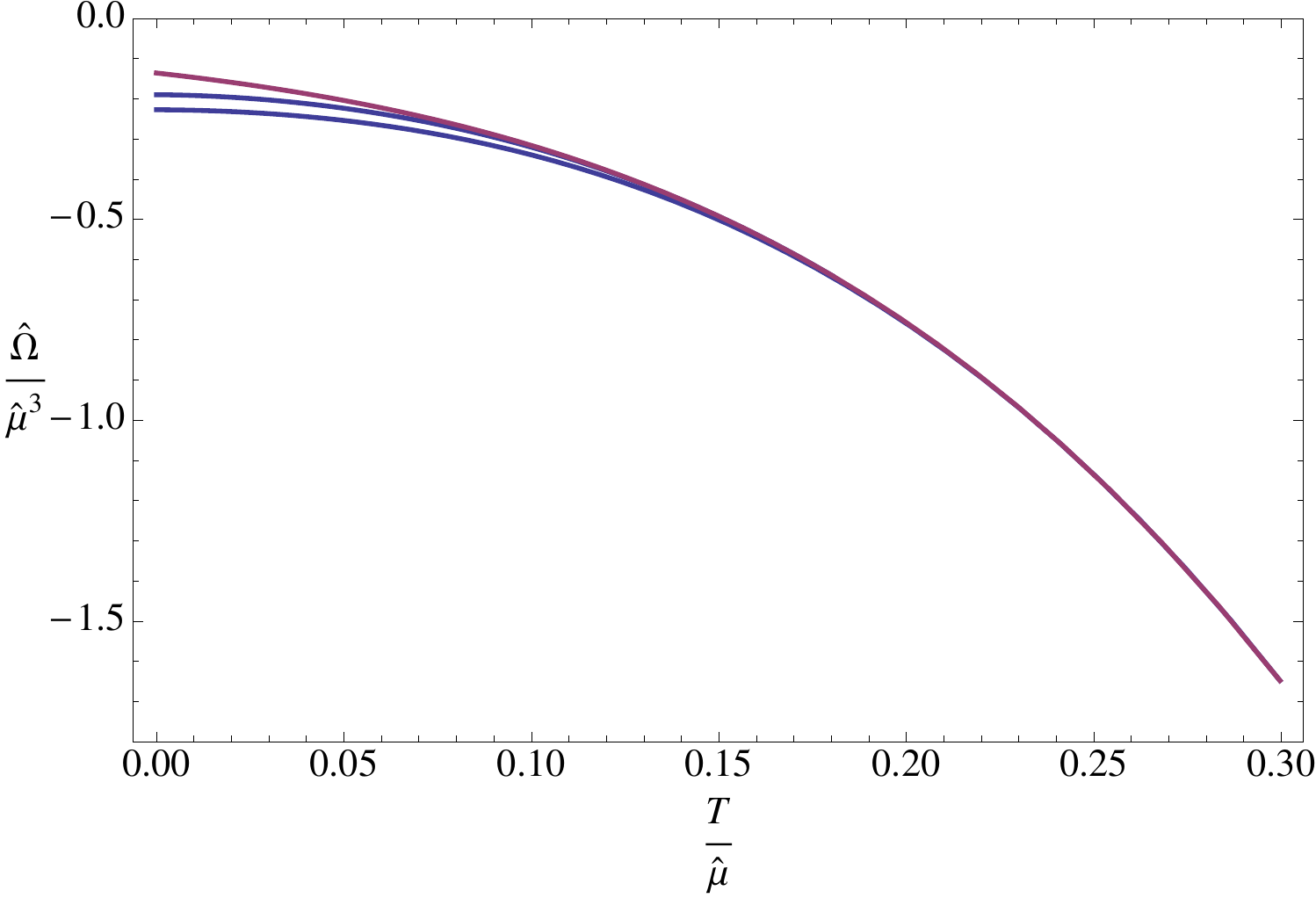}\caption{Free energy of the Reissner-Nordstrom black hole (top, red) and the free energy of two electron stars (blue) as a function of temperature. The electrons stars have $\hat \b = 10$ and $\hat m = 0.2$ (lower) and $\hat m =0.3$ (upper). The maximal temperature plotted is the critical temperature of the lower electron star. \label{fig:freeenergy}}
\end{center}
\end{figure}

Figure \ref{fig:freeenergy} compares the free energy of two electron stars with the Reissner-Nordstrom black hole in the absence of a fermion fluid. The first important observation is that the free energy of the stars is indeed lower than that of the black hole for all temperature below the transition temperature at which the star is born.
Secondly, we see that the transition is extremely soft. In the plot one can only distinguish the free energy of the black hole with and without the fermion fluid at temperatures below around half the transition temperature. Increasing the accuracy of the numerics and fitting the curves close to the transition point, we find that the transition is third order, so that $\Delta \Omega \sim (T_C-T)^3$. This fact can be understood analytically as follows.

Consider a fermion with mass $\hat m < 1$. We noted in figure \ref{fig:critical} that a fermionic fluid will start to condense at a radius $r=r_c$ and temperature $T=T_C$ given by the conditions (\ref{eq:conditions}). Cooling below the critical temperature, we saw in figure \ref{fig:buildup} how the fermion fluid spreads out as the temperature is lowered. At any temperature the fluid has width $\Delta r = r_2 - r_1$. Solving the conditions $\mu_\text{loc.}(r_{1,2}) = m$ perturbatively at small $T-T_C$, using the background Reissner-Nordstrom geometry, gives to leading order
\be\label{eq:Deltar}
\frac{\Delta r}{r_+} = \# \, (1 - T/T_C)^{1/2} + \cdots \,.
\ee
Here and below $\#$ refers to a complicated but computable mass dependent number. One can further check that to leading order in $T-T_C$ the backreaction of the fermion fluid on the geometry neither alters nor contributes to the result (\ref{eq:Deltar}). Because the backreaction of the fluid on the black hole solution appears to be subleading, we might suspect that the contribution of the fluid to the free energy is simply, from the Lagrangian (\ref{eq:lagrangian}),
\be\label{eq:cubic}
\Delta \hat \Omega = T \Delta \hat S_E = - \int_{r_1}^{r_2} \hat p \, \sqrt{f g}/r^2  dr  = - \# \, \hat \b \, T_C^3 \, (1 - T/T_C)^3 \,.
\ee
Here $\hat S_E$ is the rescaled Euclidean action, divided by the volume of the field theory spatial directions. The scaling follows immediately from (\ref{eq:Deltar}) together with the facts that, from (\ref{eq:eos}), we have $\hat p \sim \delta \hat \mu_\text{loc.}^{5/2}$ for $\hat \mu_\text{loc.} = \hat m + \d \hat \mu_\text{loc.}$ and that $\delta \hat \mu_\text{loc.} = \ocal\left( (\Delta r)^2 \right)$. This second statement in turn follows from the fact that the local chemical potential may be approximated, again simply using the Reissner-Nordstrom geometry, by the parabola $\hat \mu_\text{loc.} = \hat m + \# (1 - T/T_C)^{1/2} \D r/r_+ - \# (\D r/r_+)^2$ just below the transition temperature. To determine the coefficient in (\ref{eq:cubic}), one should use the expression (\ref{eq:eos}) for the pressure together with the parabolic expression for $\hat \mu_\text{loc.}$ and perform the integral. This can be done either numerically or analytically. Some illustrative values for the coefficient $\#$ in (\ref{eq:cubic}) are
\be
\begin{array}{c|c|c|c}
\hat m = 0.1 & \hat m = 0.3 & \hat m = 0.5 & \hat m = 0.8 \\
 \# = 0.0139 & \# = 0.7489  & \# = 3.529 & \# = 27.02
\end{array} \,.
\ee
We have checked that these values agree to within around one part in hundred with the values obtained by fitting the output of a full numerical integration of the equations of motion (using \textsc{Mathematica}'s {\tt NDSolve} with {\tt WorkingPrecision} set to 30). This provides good evidence that (\ref{eq:cubic}) is in fact the exact leading order result for the free energy difference close to the transition temperature and that there is therefore a third order phase transition in the system.

The reason the transition is so soft is a combination of two effects: immediately below the transition temperature there is a low density of fluid and it is furthermore constrained to a thin band of radii in space. While the local charge and energy densities both scale like $\delta \hat \mu_\text{loc.}^{3/2}$ close to the critical temperature, the pressure scales like $\delta \hat \mu_\text{loc.}^{5/2}$. This additional suppression feeds through to the fluid action and hence the free energy.

\begin{figure}[h]
\begin{center}
\includegraphics[height=250pt]{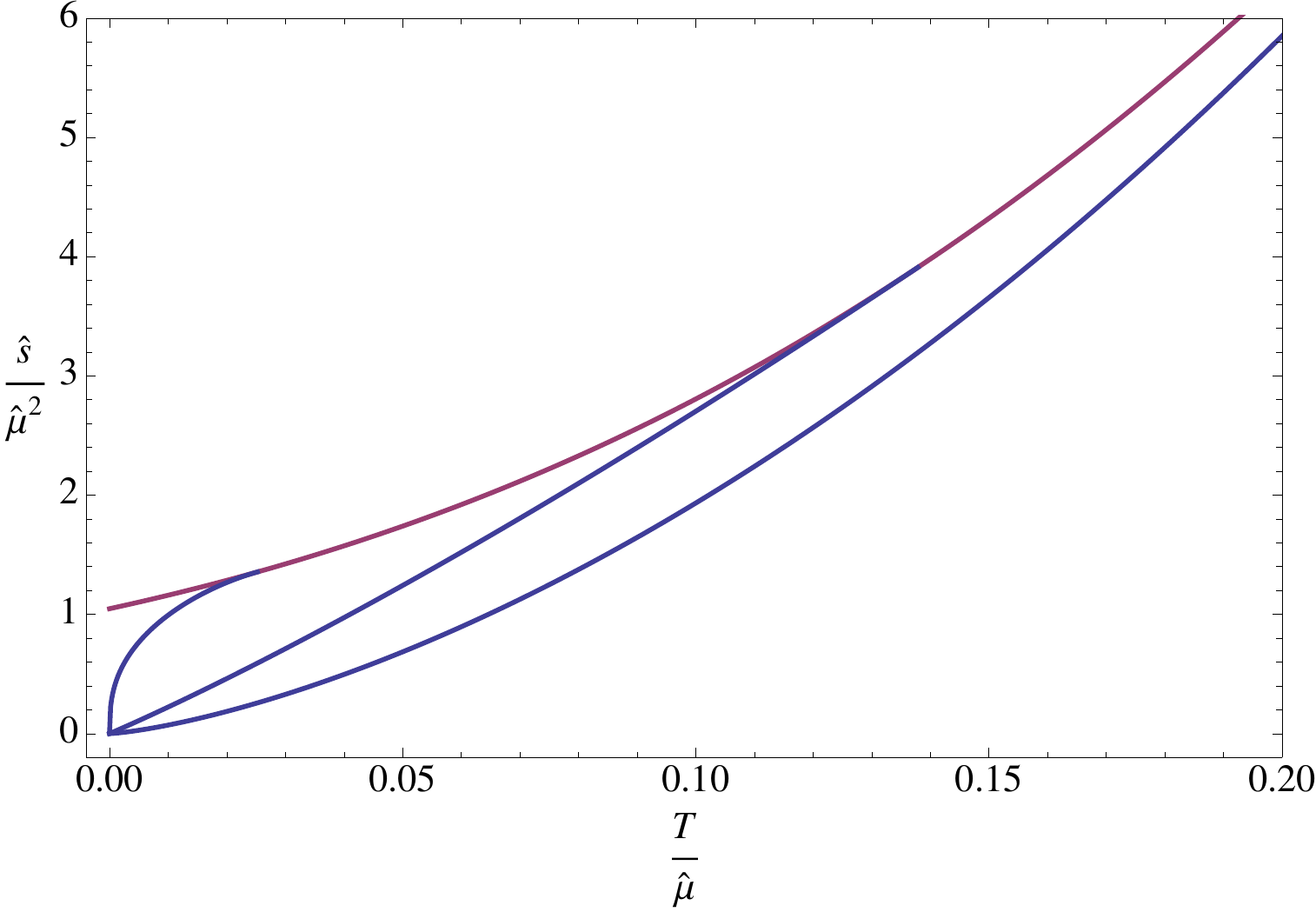}\caption{Temperature dependence of the entropy density. Red line (top) is Reissner-Nordstrom. The three blue lines (lower) are electron stars with $\hat \b = 20$ and, from left to right, $\hat m = 0.7, 0.36, 0.1$. Fitting to a power law at low temperatures leads to $z \approx 5.4, 2 ,1.5 $ respectively. \label{fig:entropy}}
\end{center}
\end{figure}

Beyond the free energy, a second interesting thermodynamic observable is the entropy density. The original motivation to consider the backreaction of fermions on black hole spacetimes was to replace the emergent IR scaling of extremal black holes, which has an infinite dynamical critical exponent $z=\infty$, with a finite $z$ scaling geometry at zero temperature \cite{Hartnoll:2009ns}. Dimensional analysis implies that at low temperatures, $T \ll \hat \mu$, the entropy density scales as $\hat s \sim T^{2/z}$. This scaling has previously been exhibited in a different holographic system in \cite{Goldstein:2009cv}. In \cite{Hartnoll:2010gu} the exponent $z$ was computed in terms of $\{\hat \b, \hat m \}$. Reading off an entropy scaling from our numerics therefore gives a vivid diagnostic of low temperature criticality as well as a check on our numerics. Figure \ref{fig:entropy} shows the entropy as a function of temperature for the black hole in the absence of an electron fluid together
with three electron stars. The entropy density of the black hole tends to a constant at zero temperature while the electron star entropy densities tend to zero. By fitting the numerical output to a power law, we can read off the dynamical critical exponent via $s \sim T^{2/z}$. The values obtained, quoted in the caption of figure \ref{fig:entropy}, agree with the zero temperature results of \cite{Hartnoll:2010gu} to within one part in a hundred.

Two issues remain to be discussed. Firstly, we promised a more detailed explanation of why it is consistent to treat a zero temperature fermion fluid in a black hole background. The black hole implies that the Euclidean time circle is periodically identified with radius $1/T$. The local charge density in (\ref{eq:eos}) should therefore necessarily include thermal excitations
\be
\sigma = 2 \pi^2 \b \sum_\pm \int \frac{d^3p}{(2\pi)^3} \frac{\pm 1}{1 + e^{(E \mp \mu_\text{loc.})/T_\text{loc.}}} \,.
\ee
The local chemical potential is again (\ref{eq:mulocal}). The local temperature $T_\text{loc.} = T/\sqrt{g_{tt}} = T/L\sqrt{f}$. It is then clear from the definitions and our limit (\ref{eq:sim}) that
\be\label{eq:tovermu}
\frac{T_\text{loc.}}{\mu_\text{loc.}} = \ocal\left(\frac{\k}{e L}\right) \ll 1 \,.
\ee
This justifies treating the local fermion fluid at zero temperature as we have done throughout. The physical content of this approximation is that while fermions are inherently quantum mechanical, the large local chemical potential allows the fermions to be treated in terms of local classical fields, and therefore suppresses the effect of Hawking radiation. As usual, the relation (\ref{eq:tovermu}) will break down parametrically close to the black hole horizon due to the diverging local temperature. A proper treatment of the quantum stress tensor will indicate that there is no physical divergence at the horizon \cite{Birrell:1982ix}.

Finally, while the low temperature phase can be characterised in the bulk by the fact that a nonzero fraction of the total charge is carried by a fermion fluid, as seen in figure \ref{fig:Qratio}, we have not discussed an order parameter of the boundary quantum field theory that could distinguish the two phases. The bulk microscopic fermion field is dual to a `single trace' fermionic operator $\Psi$ in the dual quantum field theory. It is useful to Fourier decompose this operator into creation operators $c^\dagger_{\w,k}$ \cite{deBoer:2009wk, Arsiwalla:2010bt, Hartnoll:2010xj}.
On shell bulk fermionic states have $\w = \w_k(M)$, where $M$ is a label that depends on the radial location of the bulk fermion as well as its momentum in the radial direction. Details are given in \cite{Hartnoll:2010xj} where it is also (implicitly) noted that the charge carried by fermions can be expressed as the expectation value of the `double trace' operator
\be
Q_\text{fermion.} = \int dM A(M) \int d^2k \left\langle c^\dagger_{\w_k(M),k} c_{\w_k(M),k} \right\rangle \,.
\ee
The function $A(M)$ is to be read off from the distribution of fermions in the bulk. This quantity appears not to be easy to express purely in terms of boundary quantities. A more natural boundary quantity is the generalized fermion density
\be\label{eq:n}
n = \int \frac{d\w\, d^2k}{(2\pi)^3} \left\langle c^\dagger_{\w,k} c_{\w,k} \right\rangle \,.
\ee
This expectation value is computed as follows. Each bulk fermionic state corresponds to a solution of the Dirac equation in the background electron star spacetime. Although within our Thomas-Fermi description the local fermion density drops to zero at the outer boundary of the electron star, each occupied state will have a nonvanishing tail that reaches the boundary of the spacetime.\footnote{We are grateful to Koenraad Schalm for emphasizing the importance of the tail of the Dirac equation.} Squaring the coefficient of each $\{\w,k\}$ mode will determine the expectation values in (\ref{eq:n}) via the usual holographic dictionary. It should be possible to perform this computation explicitly and we hope to return to this point in the future.

To leading order in the bulk semiclassical limit (\ref{eq:sim}), the generalized fermion density (\ref{eq:n}) acts as a field theory order parameter; it is nonzero in the low temperature phase but vanishes above $T_C$. Away from the strict semiclassical limit it is no longer an order parameter. Thermal excitations imply that the generalized density is never exactly zero. The `order parameter' does not break any symmetries of the theory. A third order transition is somewhat exotic, and deserves a better field theoretic understanding.

\vspace{0.8cm}

{\it Acknowledgements}. It is a pleasure to acknowledge very helpful discussions with Dionysios Anninos, Clay Cordova, Frederik Denef, Diego Hofman, Koenraad Schalm and Alireza Tavanfar. This research was partially supported by DOE grant DE-FG02-91ER40654, the FQXi foundation and  the Center for the Fundamental Laws of Nature at Harvard University.

\vspace{0.2cm}

{\it Noted Added}. Yesterday the paper \cite{Puletti:2010de} appeared which has a strong overlap with our results. While mostly in agreement, that paper claims the transition is second rather than third order. We believe this is because they have neglected a $-sT$ term in the free energy.

\end{document}